\documentstyle[12pt]{article}

\begin{document}
\begin{titlepage}
\title{BFV QUANTIZATION OF RELATIVISTIC SPINNING PARTICLES WITH A SINGLE
BOSONIC CONSTRAINT}
\author{Silvio J. Rabello\thanks {e-mail: goldegol@vms1.nce.ufrj.br}
\, and Arvind N. Vaidya \\
\\{\it Instituto de F{\'\i}sica}\\
{\it Universidade Federal do Rio de Janeiro}\\
{\it Rio de Janeiro  RJ}\\
{\it Caixa Postal 68.528-CEP 21945-970}\\
{\it Brasil}}
\maketitle
\begin{abstract}

{\sl Using the BFV approach we quantize a pseudoclassical
model of the spin one half relativistic particle that contains a single
bosonic constraint, contrary to the usual locally supersymmetric models
that display first and second class constraints.}

\end{abstract}
\vfill\hfill\LaTeX
\thispagestyle{empty}
\end{titlepage}

\setcounter{page}{1}

The dynamics of the relativistic spinless particle, of mass m
and 4-momentum $P^\mu$ is generated by the vanishing canonical Hamiltonian
$H_0=P^2-m^2\approx 0$ \cite{HeTe2},  reflecting the invariance under
reparametrizations of the world line $X^\mu(s)$ ($s\in [0,1]$) .
To describe spinning relativistic particles one often promotes the
reparametrization invariance of the spinless action to a local supersymmetry
on s \cite{Brink} (one dimensional supergravity). The resulting
pseudoclassical system has first and second class constraints that after an
appropriate treatment (e.g. the Dirac algorithm) reduce to two first class
constraints, one bosonic the other fermionic, the mass-shell condition and
its ``square root'' respectively. In this paper we study another
description of a spinning particle, where only one bosonic constraint is
needed. The inspiration for this model comes from the original Dirac
Hamiltonian \cite{Di} \footnote{$\hbar=c=1$ and
$\eta_{\mu\nu}=diag(1,-1,-1,-1)$}:
\begin{equation}
\label{Dirac}
H_{D}=\gamma_0\mbox{\boldmath $\gamma\cdot P$} +\gamma_0 m
\end{equation}
where ${\bf P}$ is the 3-momentum and the $\gamma '$s are the Dirac
matrices. The above $H_{D}$ generates translations in the physical time
$X^0$ and it is not a Lorentz scalar. In order to have a model for the
relativistic spinning particle in the same lines of the spinless case we
study an analog of (\ref{Dirac}) that is Lorentz covariant, generate
translations in the parameter time s and is weakly zero so that the
resulting theory is invariant under reparametrizations of $X^\mu(s)$\,.
Introducing the commuting Lagrange multiplier $\lambda$ and anticommuting
coordinates $\zeta_5$ and $\zeta_\mu$ $(\mu=0,1,2,3)$ , the pseudoclassical
Lagrangian reads:
\begin{equation}
\label{Lag}
{\cal L}=P^\mu\dot X_\mu + {i\over 2}(\zeta_5\dot\zeta_5 -
\zeta^\mu\dot\zeta_\mu) - \lambda H_0\,,
\end{equation}
with the canonical Hamiltonian given by
\begin{equation}
\label{H}
H_0=\zeta_5\zeta_\mu P^\mu -m \approx 0 \,,
\end{equation}

{}From this Lagrangian we can read the sympletic structure that on
quantization gives the (anti)commutation relations
(the non zero part of it) :
\begin{equation}
\label {commu}
[X^\mu,P^\nu]_{-}=i\eta^{\mu\nu}\,,\qquad [\zeta_{\mu},\zeta_{\nu}]_{+}
=-2\eta_{\mu\nu}\,,\qquad [\zeta_5,\zeta_5]_{+}=2 \,.
\end{equation}

In the quantum mechanics of this model the constraint (\ref{H}) becomes a
gauge generator, that in the Dirac theory of constrained systems \cite{HeTe}
annihilates the physical states, i.e. $H_0\vert phys\rangle = 0$. For
consistency we have that its square,
\begin{equation}
\label{Constr2}
H_0^2=P^2 + m^2 - 2m\zeta_5\zeta_\mu P^\mu\,,
\end{equation}
must also annihilate $\vert phys\rangle $. It is easy to see that on
physical states we have :

\begin{equation}
\label{m-shell}
H_0^2\vert phys\rangle = (P^2-m^2)\vert phys\rangle =0\,.
\end{equation}
Thus our single bosonic constraint (\ref {H}) implies the mass-shell
condition of a relativistic particle of mass m. We now proceed to the
BFV quantization of the model.

In the BFV formulation \cite {BV,HeTe} the hermitian nilpotent BRST
charge operator for this model is given by
\begin{equation}
\label{charge}
Q=\eta H_0 +\bar{\cal P}\pi\,,
\end{equation}
where we extended our phase space by introducing the conjugate momentum
$\pi$ for the Lagrange multiplier $\lambda$, a fermionic  pair of conjugate
ghost operators $(\eta\,,\,{\cal P})$ for the constraint $H_0$,
and a pair of fermionic antighost conjugate operators
$({\bar\eta}\,,\,{\bar{\cal P}})$ to take care of the constraint
$\pi$ that appear due to the ``einbein'' character of $\lambda$
that we assume from start. These hermitian operators generate the algebra
(the non zero part of it)
\begin{equation}
\label{GhostA}
[\lambda,\pi]_-=i\,,\;\;\;[\eta, {\cal P}]_{+}=
[\bar{\eta}, \bar{\cal P}]_{+}=1\,.
\end{equation}

Due to the nilpotency of Q the following extension of
the original BRST invariant Hamiltonian $H_0$ is also invariant :
\begin{equation}
\label {ext}
H_0\rightarrow H_0+[\Psi,Q]_{+}\,.
\end{equation}
Where $\Psi$ is an arbitrary ``gauge fermion'', that we here choose to be
$\Psi={\cal P}\lambda$. And also, since $Q\vert phys\rangle=0$, the above
extension of $H_0$ can be used in the evolution operator without changing
the transition amplitudes. We now consider the transition amplitude
\begin{equation}
\label{K}
K(z'',z';s)\equiv \langle z''\vert e^{-iHs}\vert z'\rangle
=\langle z'',s\vert z',0\rangle\,,
\end{equation}
in the basis $\vert z';s\rangle \equiv \vert P',\lambda',\eta',{\bar\eta}';s
\rangle $ with $\langle z'',0\vert z',0\rangle=\delta (z''-z')$
(hereafter all operator eigenvalues will be primed). The above $K(z'',z';s)$
obeys the Schr\"odinger equation in the parameter time s:
\begin{equation}
\label {Sch}
i{\partial K(z'',z';s)\over{\partial s}}=\langle z'',s\vert H\vert z',0
\rangle\,,
\end{equation}
with $H$ being
\begin{equation}
\label{Ham}
 H=[\Psi,Q]_{+}=\lambda(\zeta_5\zeta_\mu P^\mu -m)+i{{\cal P}}
 {{\bar{\cal P}}}\,.
\end{equation}
The expectation value  $\langle z'',s\vert H\vert z',0\rangle $
is easily evaluated if we apply a method developed long ago by Schwinger
\cite{Schw} that consists in solving the Heisenberg equations for the
canonical variables, writing the ${\cal P}$ and $\bar{\cal P}$ operators in
terms of the $\eta$ and $\bar\eta$ operators and inserting these in a time
ordered fashion in $H$. After that we can easily integrate (\ref{Sch}). For
this model we have the Heisenberg equations (the dot means derivative with
respect to s) :
\begin{eqnarray}
\label{Heis2}
\dot X^\mu &=& \lambda\zeta_5\zeta^\mu\,,\qquad\quad\;\;\;
\dot P^\mu=0\,,\\
\dot \zeta^\mu &=&-2i\lambda\zeta_5 P^\mu \,,\qquad\,
\dot\zeta_5=-2i\lambda\zeta_\mu P^\mu\,,\\
\dot\lambda &=& 0\,,\;\qquad\qquad\qquad
\dot\pi=-(\zeta_5\zeta_\mu P^\mu -m)\,,\\ \dot\eta &=& {\bar{\cal P}}\,,
\qquad\qquad\qquad \dot{\cal P}=0\,,\\
\dot{\bar\eta} &=& -{\cal P}\,,\;\quad\qquad\qquad  \dot {\bar{\cal P}}=0\,.
\end{eqnarray}

As in the case of the spinless particle \cite{HeTe2} the above equations
carry an ambiguity, take the expectation value of $\dot\zeta_\mu$, if we
change $\lambda\rightarrow -\lambda$ and $P_\mu\rightarrow -P_\mu$ we return
to the same equation. A remedy to solve this ambiguity in expectation values,
is to restrain the eigenvalues of $\lambda$ to be either positive or negative
\cite{HeTe2}.

The solutions for the ghosts and antighosts are:
\begin{equation}
\label{Sol1}
\eta(s)=\eta(0)+{\bar{\cal P}}s\,,\qquad\qquad
{\bar\eta}(s)={\bar\eta}(0)-{\cal P}s\,.
\end{equation}

We now write  ${{\cal P}}$ and ${\bar{\cal P}}$ in terms of ${\eta}(s)
,\;{\eta}(0),\;{\bar{\eta}}(s)$ and ${\bar{\eta}}(0)$. For the $\zeta '$s
we take as initial values $\zeta_5 (0)=\gamma_5$ and $\zeta_5(0)\zeta_\mu(0)
=\gamma_\mu$\,. Since $H$ is s-independent we from now on substitute these
values in it, so that we get the time ordered Hamiltonian :
\begin{eqnarray}
\label {Hord1}
H_{ord}&=&\lambda(\gamma_\mu P^\mu-m)-
{i\over s^2}\biggl({\bar\eta}(s)\eta(s)-
{\bar\eta}(s)\eta(0)\nonumber\\
&+&{\bar\eta}(0)\eta(0)+\eta(s){\bar\eta}(0)
-[{\bar\eta}(0),\eta(s)]_{+}\biggr)\,,
\end{eqnarray}
with $[{{\bar\eta}}(0),{\eta}(s)]_{+}=s$\,. We are now in position to
integrate (\ref{Sch}):
\begin{equation}
\label {W1}
ln\,K=-is\lambda'(\gamma^\mu P'_\mu-m)-{1\over s}({\bar\eta}''-{\bar\eta}')
(\eta ''-\eta') -ln\,s +\Phi\,,
\end{equation}
where $\Phi$ is a s independent function of the dynamical variables.
With aid of the boundary condition $\langle z'',0\vert z',0\rangle=
\delta (z''-z')$ :
\begin{equation}
\label {Phi2}
\Phi=ln\,i\delta^4(P''-P')\delta(\lambda''-\lambda')\,.
\end{equation}

Now that we have found $K$ we must impose the invariance under Q, i.e.
Q acting each argument of $K(z'',z';s)$ must produce zero as a result. Among
the several ways for this to be true \cite {HeTe}, we choose
the boundary conditions :
\begin{equation}
\label {Bound}
\pi''=\pi'=\eta''=\eta'={\bar\eta''}={\bar\eta'}=0\,.
\end{equation}
The condition on ${\bar\eta}$ is a consistency one for $\pi=[Q,
{\bar\eta}]_{+}$ and $Q$ to annihilate $\vert phys\rangle$. We now Fourier
transform $K$ on $\lambda''$ and $\lambda'$ choosing $\lambda\geq0$,
\begin{equation}
\label {SchwRep1}
K=i \int_0^{\infty} d\lambda'\; s e^{i\lambda'(\pi''-\pi')}
e^{-i[s\lambda'(\gamma^\mu P'_\mu-m)+{i\over s}
({\bar\eta}''-{\bar\eta}')(\eta ''-\eta')]}\delta^4(P''-P')\,.
\end{equation}
With the BRST invariant boundary conditions and defining $s\lambda'\equiv T$ :
\begin{equation}
\label {SchwRep2}
K=i\int_{0}^{\infty}dT\; e^{-iT(\gamma^\mu P'_\mu-m)}
\delta^4(P''-P')={\delta^4(P''-P')\over{\gamma^\mu P'_\mu -m -i0_{+}}}\,.
\end{equation}
The choice of positive $\lambda$ led to the momentum space
Feynman propagator for Dirac fermions , had we chosen negative values for
$\lambda$ we would end up with the complex conjugate of it.
\bigskip

\noindent{\Large\bf Discussion}

\smallskip
It would be interesting to introduce interactions with background fields
as well as to generalize our Lagrangian to describe extended objects like
spinning strings and membranes. This work was partially supported
by the CNPq.


\begin{thebibliography}{99}


\bibitem{HeTe2} M. Henneaux and C. Teitelboim  {\it Ann. Phys.}
{\bf 143}, 127 (1982).

\bibitem{Brink} L. Brink , P. Di Vecchia  and P. Howe {\it Nucl. Phys.} B
{\bf 118}, 76 (1977).

\bibitem{Di} P.A.M. Dirac {\it The Principles of Quantum Mechanics}
(3rd edn) (Oxford University Press, 1947).

\bibitem{HeTe}M. Henneaux  and C. Teitelboim  {\it Quantization of
               Gauge Systems} (Princeton Univ. Press, 1992);
               M. Henneaux  {\it Phys. Rep.} {\bf 126}, 1 (1985).

\bibitem{BV} E.S. Fradkin and  G.A. Vilkovisky  {\it Phys. Lett.}  {\bf 55B},
224 (1975); I.A. Batalin and G.A. Vilkovisky {\it Phys. Lett.}
{\bf69B}, 309 (1977).

\bibitem{Schw}J. Schwinger {\it Phys. Rev.} {\bf 82}, 664 (1951).


\end{thebibliography}
\end{document}